\documentstyle[aps,preprint,psfig]{revtex}
\begin{document}
\draft

\title{Correlations of conductance peaks and transmission phases in 
deformed quantum dots} 
\author{Reinhard Baltin$^1$, Yuval Gefen$^2$, Gregor Hackenbroich$^{1,3}$,
  and Hans A. Weidenm\"uller$^1$} 
\address{$^1$Max Planck Institut f\"ur Kernphysik,
  Postfach 103980, 69029 Heidelberg, Germany} 
\address{$^2$Department of Condensed Matter Physics,\\ The Weizmann Institute
  of Science, 76100 Rehovot, Israel}
\address{$^3$Universit\"at GH Essen,
  Fachbereich 7, 45117 Essen, Germany} 
\date{\today} 
\maketitle
\begin{abstract}
  We investigate the Coulomb blockade resonances and the phase of the
  transmission amplitude of a deformed ballistic quantum dot weakly
  coupled to leads. We show that preferred single--particle levels
  exist which stay close to the Fermi energy for a wide range of
  values of the gate voltage. These states give rise to sequences 
  of Coulomb blockade resonances with correlated peak heights and
  transmission phases. The correlation of the peak heights becomes
  stronger with increasing temperature. The phase of the transmission
  amplitude shows lapses by $\pi$ between the resonances. Implications
  for recent experiments on ballistic quantum dots are discussed.
\end{abstract}

\section{Introduction}
Quantum dots have been intensively investigated both experimentally
and theoretically \cite{kou} in recent years. In this paper, we
present a theoretical study of the correlations of conductance peaks
and of transmission phases that have been observed in recent
experiments on quantum dots in the Coulomb blockade regime.

Quantum dots are small islands of electrons that are only a few
hundred nanometers in size and typically contain a few hundred
electrons. The spectrum of a quantum dot is determined by the Coulomb
interaction of the electrons and by the external electrostatic
confining potential. The confining potential and hence the size and
shape of a quantum dot can be controlled by external gates. This makes
quantum dots an ideal tool for studying the properties of finite
systems of interacting fermions. 

Experimentally, the spectra of quantum dots have been measured using
optical (far--infrared) spectroscopy and/or transport experiments. In
the latter case, the quantum dot is coupled via tunnel barriers to
external leads. The conductance measured at a finite drain--source
voltage reveals the excitation spectrum of the dot whereas the linear
conductance yields the addition spectrum of the quantum dot. Both the
excitation and the addition spectrum are dominated by the classical
Coulomb blockade effect: Large conductance peaks are observed when the
dot potential is tuned in such a way that the number of electrons on
the dot can fluctuate without any cost in energy. These peaks are
nearly periodic in the gate voltage on the dot. At consecutive peaks
the number of electrons on the dot changes by one. At values of the
gate voltage located between the positions of the conductance peaks,
electron transport through the dot requires a large charging energy. 
Hence, the current between conductance peaks is strongly suppressed,
the remaining current being mostly due to virtual tunneling processes
(co--tunneling regime) \cite{avnaz}.

Typically, metallic quantum dots have such a large density of states 
that the Coulomb
blockade oscillations can be described by classical theory which
ignores the discreteness of the spectrum. The situation is different
in semiconductor dots. Here, the mean single--particle level spacing
$\Delta$ can be larger than the temperature $k T$. The regime $\Delta
\gg k T, \Gamma $, where $\Gamma$ is the strength of the coupling to
the leads, is the resonant tunneling regime. In this regime, each
conductance peak is mediated by a single quantum state of the dot. The
peak height of the conductance resonance is a direct signature of the
wave function of the resonant state.

Some years ago, Jalabert, Stone, and Alhassid developed a statistical
theory of the Coulomb blockade \cite{jalabert} in the resonant
tunneling regime. In order to explain
strong fluctuations of the peak heights of neighboring conductance
resonances, they assumed that the eigenstates of a quantum dot can be 
described by random matrix theory. According to this theory no
correlations are to be expected for the peak heights of neighboring
peaks. Moreover, the distribution of peak heights is predicted to be
universal, and only determined by the fundamental symmetries of the
system. 
While two recent experiments \cite{chang,marcus} appeared to have
confirmed these predictions, a more careful examination of the
experimental data casts some doubt on the validity of random matrix
theory in describing the physics of quantum dots. Deviations from the
predictions of random matrix theory include the appearance of 4-5
correlated transmission peaks \cite{marcus}, non Wigner-Dyson related
distribution of conductance peak spacing \cite{sivan,wharam},
and reduced sensitivity to magnetic flux
\cite{chang,marcus,wharam,berko}. 
Even stronger correlations have
been reported in very recent experiments using a quantum dot embedded
in an Aharonov-Bohm ring \cite{heiblum} where both the peak heights
and the phase of the transmission amplitude could be measured. Strong
correlations within sequences of more than 10 resonances were found.

In the experiments of
Refs.~\cite{chang,marcus,wharam,heiblum}, $k T$ was approximately
$(0.1-0.5) \Delta$.
When the resonant tunneling limit $\Delta\gg kT$ is not fully met, 
also neighboring eigenstates around the Fermi energy (rather than only
a single state) contribute to the conductance peak. 
Extending the random matrix theory
description to the regime $\Delta > k T$, correlations in the peak
height were found \cite{pcalhassid}. 
Those temperature-induced correlations seem sufficiently strong to
account for sequences of up to 5 correlated peaks as reported in
Ref.~\cite{marcus}. However, temperature alone can clearly be ruled
out as the source of the much stronger correlations found in
Ref.~\cite{heiblum}. A mechanism different from and acting in addition
to temperature must be the origin of these correlations.

In this paper we propose such a mechanism. 
Our model involves certain geometry-specific assumptions and is
therefore restricted in generality and universality. Nevertheless it
may still pertain to experiments on nearly integrable ballistic dots
(cf. Section \ref{CondOsc}). 
Our mechanism is a synthesis of two
approaches developed earlier \cite{hhw,yy}. In Ref.~\cite{hhw} a
scenario was described for peak correlations at {\em vanishing
  temperature}. It was argued that deformation of the confining
potential of the dot generically gives rise to avoided crossings of
the single--particle levels on the dot. As a result of such crossings
one and the same eigenstate of the dot may dominate a sequence of
neighboring conductance peaks \cite{hhw} and thereby cause
correlations. Although several arguments in support of this mechanism
were given \cite{hhw}, a quantitative study of the resulting
correlations has not been presented yet. A different line of thought
was pursued in Ref.~\cite{yy}. That paper aimed at explaining the
"phase lapse" observed in the experiment of Ref.~\cite{heiblum} (a
different theoretical discussion of the phase lapse was presented in
Ref.~\cite{hw}). It was shown \cite{yy} that the phase--lapse behavior
as well as strong {\em finite temperature} correlations of conductance
peaks arise if the dot supports one particularly well conducting
state. This state would dominate a sequence of conductance peaks
provided it remained within an energy interval of order $kT$ around the
Fermi energy.

Here we study peak--height correlations and the transmission
phase in quantum dots taking a deformed harmonic oscillator as a
specific example. Using this model, we show that deformation of the
confining potential leads to peak--height correlations, in keeping
with the arguments of Ref.~\cite{hhw}. We identify those eigenstates
of the quantum dot that are most strongly coupled to the external
leads and which, therefore, support the bulk of the current through
the dot. Peak--height correlations are strongest for sequences of
resonances mediated by such states. We also investigate the influence
of temperature on the correlations. As expected from the studies in
Ref.~\cite{yy}, we find that temperature leads to a marked increase of
the correlations. Combining the effect of deformation and of temperature,
we obtain sequences of up to 30 conductance resonances with similar
transmission phases and similar peak heights. We demonstrate that this
same mechanism may also account for the phase--lapse behavior
observed in Ref.~\cite{heiblum}.

The paper is organized as follows: In the next Section we introduce a
model for a deformed quantum dot and describe how correlations can
arise. In Section 3 we calculate the conductance with the help of a
master equation. In Section 4 we investigate the transmission
phase. This phase can be measured in Aharonov--Bohm type experiments
containing a quantum dot. The last Section gives a summary and a
discussion of the limitations of our model.

\section{The Model}
\label{Model}
The confining potential of a quantum dot is often
defined electrostatically in terms of a split gate. As depicted in
Figure~1, this is an arrangement of electrodes on the surface of a
heterostructure. When a negative bias is applied to the gates, the
two--dimensional electron gas located some 100 nm or so beneath the
electrodes will be depleted. The barriers through which electrons can
tunnel between the leads and the dot are denoted by $A$ and $B$. A
voltage $V_g$ applied to the plunger gate $P$ controls the chemical
potential on the dot. A change of $V_g$ not only changes the number of
electrons on the dot but also distorts the confining potential of the
two--dimensional electron gas in a substantial way, causing a
deformation of the quantum dot \cite{hhw}.

We consider the standard Hamiltonian $H$ for a quantum dot coupled to
leads, containing the Hamiltonians $H^{L}$ and $H^{R}$ of the left and
right leads, respectively, the Hamiltonian of the isolated quantum dot
$H^D$, and the Hamiltonian $H^T$ for tunneling between the leads and
the dot, 
\begin{eqnarray}
  H&=& H^{L} + H^{R} + H^D + H^T \ , \\
  H^{L(R)} &=& \sum_{k} \epsilon^{L(R)}_k \ {a^{L(R)}_k}^{\dagger}
  a^{L(R)}_k \ , \nonumber \\
  H^D &=& \sum_{n} \epsilon_n c^{\dagger}_n c_n + \frac{1}{2} U
  \hat{N} (\hat{N} -1) \ , \nonumber \\
  H^T &=& \sum_{n,k} \left( V^L_{n,k} a^L_k c^{\dagger}_n +
  \mbox{h.c.} \right) + \sum_{n,k} \left( V^R_{n,k} a^R_k
  c^{\dagger}_n + \mbox{h.c.} \right)\ \ . \nonumber
\end{eqnarray}
Here, $\epsilon_k^{L,R}$ and $\epsilon_n$ are the energies and 
$a_k^{L,R}$ and $c_n$ the annihilation operators for single--particle
states in the leads and in the dot, respectively. For the Coulomb
interaction on the quantum dot we use the constant interaction model
with $\hat{N}=\sum_n n_n$ the number of electrons on the quantum dot. 
The tunneling matrix elements $V^{L(R)}_{i,k}$ involve the overlap of
wave functions in the leads and in the dot and are given below. 

We model the confining potential as an anisotropic harmonic oscillator
potential. A harmonic potential has been used previously in studies of 
quantum dots \cite{para} and, at least for small
dots, is believed to be a fair approximation to the true confining
potential. Although we use a specific model, most of our conclusions
apply to any sufficiently smooth confining potential for which the
Hamiltonian is nearly integrable. Then, the transverse and
longitudinal modes in the dot are nearly decoupled, cf. Eq.~(\ref{spec}). 
This condition is met when the matrix elements of the perturbation
violating integrability are smaller than the mean single--particle
level spacing. However, even mild disorder or boundary roughness
violating this condition will modify our picture considerably.

The energy eigenvalues $\epsilon_n$ for the quantum numbers $n =
(n_x,n_y)$ are given by
\begin{eqnarray}
\label{spec}
\epsilon_n=E(n_x,n_y) &=& \hbar \omega_x (n_x+\frac{1}{2}) + \hbar
\omega_y(V_g) 
(n_y+\frac{1}{2}) - \alpha V_g +E_0.
\end{eqnarray}
To describe the deformation, we assume that the oscillator frequency
$\omega_y(V_g) = \omega_x (1-\gamma (V_g-V_0))$ in the transverse
direction $y$ depends linearly on the gate voltage $V_g$ while 
the frequency $\omega_x$ in the longitudinal $x$--direction is held 
fixed. The parameter $\alpha$ relates the overall depth of the dot's
potential to the gate voltage. The constants $E_0$ and $V_0$ determine
the number of electrons on the dot at zero deformation.

The dependence of the single--particle levels on the gate voltage is
shown in Figure~\ref{shell}. The shell structure of the isotropic
harmonic oscillator ($V_g=V_0$) is clearly visible. It survives for
small values of $V_g$ but is eventually destroyed by deformation. Each
shell $q$ is characterized by non--negative integer quantum numbers
$q = n_x + n_y$. In each shell, there are levels depending weakly
(strongly) on $V_g$, characterized by large (small) values of $n_x$
and small (large) values of $n_y$, respectively. These are referred to
as ``flat levels'' and ``steep levels'', respectively. A small
deviation from integrability will change the level crossings shown in
Figure~\ref{shell} into avoided crossings. For nearly integrable
systems, the wave functions retain their character across avoided
crossings. Flat levels are particularly stable, their wave functions
change little with deformation (or gate voltage) and remain
self--similar even after several avoided crossings \cite{hhw}.

The matrix elements $V^L, V^R$ for tunneling from the left and right
lead to the quantum dot are given by \cite{duke}
\begin{eqnarray}
\label{tunnel}
V^{L(R)}(k,n_x,n_y) &=& \frac{\hbar^2}{2m} \int_{B} dy \left[ \psi_k(x,y)^*
  \frac{\partial \Phi_{n_x,n_y}(x,y)}{\partial x} - \Phi_{n_x,n_y}
  \frac{\partial  \psi_k (x,y)^*}{\partial x} \right]_{x=x_B} \ . 
\end{eqnarray}
Here $\psi^{L(R)}_k$ is the wave function with wave vector $k$ in the
left (right) lead , and $\Phi_{n_x,n_y}$ is the wave function in the
dot. The integration extends in the $y$--direction and $x_B$ is arbitrary
but must be located within the barrier \cite{duke}. We restrict
ourselves to the case of a single transverse channel in each lead. The
nodes of the wave functions of flat (steep) levels with large $n_x$
($n_y$) are predominantly carried by the x--component (y--component,
respectively). Thus, the wave functions of flat levels extend much
further into the barrier region and have considerably larger matrix
elements $V^{L(R)}(k,n_x,n_y)$ than those of the steep levels. This
important property is illustrated in Figure~\ref{over}. It has immediate
consequences for the conductance at finite temperature: The very same
single--particle state can dominate different Coulomb blockade
resonances seen at different values of $V_g$ \cite{hhw}. We now
explain this feature qualitatively, postponing a detailed discussion 
to later sections. 

At low temperature ($kT \ll \Delta$) and for small bias voltage, the
transmission through the dot can be qualitatively obtained from the
mean--field approximation for the dot spectrum. In this approximation,
each single--particle energy $\epsilon_i$ is replaced by the effective
value $\varepsilon_i = \epsilon_i+U \sum_{j \neq i} \langle n_j
\rangle$ \cite{anderson,hw1}. According to Koopmans' theorem,
$\varepsilon_i$ is the energy needed to add an electron in state $i$
to the dot, whereas the excitation energy at fixed electron number is
given by the difference of the corresponding two effective energies. 
Because of the Coulomb interaction, there is a gap of magnitude $U$
between the last occupied and the first empty effective
single--particle level, while the other occupied (empty) levels below
(above) the Fermi energy $E_F$ are on average separated by the usual
mean level spacing $\Delta$. Avoided crossings of single--particle
levels result in avoided crossings of the effective levels, in spite
of this gap \cite{hhw}.

A Coulomb blockade resonance occurs, and the number of electrons on the
dot changes by one, whenever an effective single--particle level crosses
the Fermi energy $E_F$ of the reservoirs. 
(We assume $E_F$ to be independent of the gate voltage in the
following). For $U \gg \Delta$, the
distance between adjacent resonances is $\delta V_g=U/\alpha$. Without
level crossings, different resonances correspond to different
single--particle levels. In the presence of level crossings, the
situation changes. This is shown in Figure~\ref{avoid} which displays
the gap between the filled levels below and the empty ones above $E_F$. 
Resonances occur at gate voltages $V_1,V_2$ and $V_3$. Suppose a flat
level $F$ (dashed) crosses $E_F$ at $V_1$. If there is an avoided
crossing of $F$ with a steep level $S_1$ from a higher shell between
$V_1$ and $V_2$, level $F$ is pushed above $E_F$ while level $S_1$ is
immersed into the Fermi sea. At $V_2$ the flat level $F$ crosses $E_F$
again, causing another resonance to occur. The mechanism works again
between $V_2$ and $V_3$ where another steep level $S_2$ intersects
with $F$. Since flat levels keep their wave functions after avoided
crossings, the resonances at $V_1, V_2$ and $V_3$ all carry the $\it
same$ single--particle wave function. This mechanism gives rise to
strong correlations of the properties (peak height and transmission 
phase) of several resonances. However, it leads to strong correlations
only if there is {\em one and only one} crossing of the flat level $F$
with a steep level within subsequent intervals $\delta V_g$. If -- on
average -- there is less than one (more than one) crossing, the level
$F$ will eventually be pulled down below (up above) the Fermi energy
and will become irrelevant for the behavior of Coulomb blockade
resonances.

So far, we have focussed attention on the resonant tunneling regime
$\Gamma, kT \ll \Delta$ where only the level $\it at$ the Fermi energy
determines the properties of the resonance. Essential modifications
arise for finite temperatures $kT \sim \Delta$. Here, also levels at a
distance $\sim kT$ from $E_F$ contribute to the resonance. Since the
flat levels are coupled to the leads much more strongly than the steep
ones, the presence of a flat level at a distance $\sim kT$ from $E_F$
suffices for it to dominate the resonance. Long sequences of
correlated resonances may occur if repeated avoided crossings cause a
flat level $F$ to stay sufficiently close to $E_F$ over a sufficiently
long range of $V_g$ values. This is the picture we investigate
quantitatively for the case of an anharmonic oscillator in the sequel.
The picture suggests that the correlations of consecutive Coulomb
blockade resonances will increase with temperature.

The number of intersection points of a flat level ($n_x \neq 0 \ ,
n_y=0$) with steep levels from higher shells can be determined from
Eq.~(\ref{spec}). The total number of crossings of the flat level
occurring in the range $V_g-V_0$ is given by 
\begin{eqnarray}
N_c &=& \frac{n_{x}^{2}}{2} \frac{\gamma (V_g-V_0)}{1-\gamma (V_g-V_0)}.
\end{eqnarray}
The number of steep levels from higher shells increases with
deformation and causes $N_c$ to increase, too, until it diverges for
the unphysical situation of extreme deformation $\omega_y \rightarrow
0$. In order to have one crossing within an interval $\delta
V_g=U/\alpha$ we require $(\partial N_c)/(\partial V_g) =
\alpha/U$. This condition yields the value of $V_g$ where maximal
correlations between Coulomb blockade resonances should occur. We
estimate the number $\Delta N$ of resonances for which the distance
between the flat level $(n_x,0)$ and the Fermi energy is less than one
level spacing and obtain
\begin{eqnarray}
\Delta N &\simeq& 2 \sqrt{n_x 
\sqrt{\frac{\alpha}{2\gamma U}}} \ ,
\end{eqnarray}
with a deformation $\omega_y/\omega_x=\sqrt{(\gamma U)/(2 \alpha)}\
n_x$.

\section{Coulomb blockade Resonances}
\label{CondOsc}

In this section we calculate the conductance $G$ of a quantum dot with
the single--particle spectrum (\ref{spec}) in linear response. We use
the master equation \cite{ben}. In the Coulomb blockade regime, maxima
in $G$ occur for values of $V_g$ where the configurations with $N$ and
$N+1$ electrons on the dot are degenerate. The resulting sharp
conductance peaks are almost equally spaced \cite{kou,ben}. For $kT
\ll \Delta$ each peak is due to a single level $n$ of the dot, and the
peak height is given by $G\sim \Gamma^L_n \Gamma^R_n / (\Gamma^L_n +
\Gamma^R_n)$ \cite{ben} with the tunneling rates $\Gamma^{L(R)}_n\sim
\sum_{k} |V_{k,n}^{L(R)}|^2 \delta (\epsilon_n-\epsilon_k^{L(R)})$. We
note that the peak heights are highly sensitive to the wave functions
in the dot. Strongly coupled levels give higher peaks than the weakly
coupled ones. For finite temperature several single--particle states
contribute to a resonance.

Under the assumption of sequential tunneling, transport through the
dot at finite temperature is described by the master equation \cite{ben}.
In the regime $kT \gg \Gamma$ this equation determines the occupation
probabilities $P_{\nu}$ of the single--particle levels of the dot
under the influence of the interaction $U$ and of the coupling to the
leads. It is given by  
\begin{eqnarray} 
\frac{\partial}{\partial t} P_{\nu} &=& \sum_{\mu (\neq \nu)} P_{\mu}
\Gamma_{in} (\mu \rightarrow \nu) - P_{\nu} \Gamma_{out} (\nu
\rightarrow \mu) \ .
\label{master}
\end{eqnarray}
Here $\mu$ and $\nu$ label Fock states, i.e., Slater determinants
defined in terms of the occupation numbers of all single--particle
states in the dot. The symbols $\Gamma_{in}$ and $\Gamma_{out}$ stand
for the rates of the tunneling processes into and out of the
dot. These processes change the number of electrons on the dot by one,
and the associated Fock states from $\mu$ to $\nu$, and vice versa. 
The rates contain not only the coupling of the specific
single--particle states of the dot to the leads but also take into
account a possible suppression of tunneling by the occupation of
the states in the leads. By expanding Eq.~(\ref{master}) around the
equilibrium probability distribution Beenakker \cite{ben} obtained the
conductance $G$ for small bias voltage,
\begin{eqnarray}
  G &=&\frac{e^2}{kT} \sum_{n} \sum_{N=1}^{\infty}
  \frac{\Gamma^L_{n} \Gamma^R_{n}}{\Gamma^L_n+ \Gamma^R_n} \ P^{\rm
  eq}_N \ [1 - F^{\rm eq}(\epsilon_n|N)] \ f(\epsilon_n + U \cdot N) \
  .
\label{condbeen}
\end{eqnarray} 
Here, $f$ is the Fermi function and $P^{\rm eq}_N$ is the probability
to find $N$ electrons on the dot,
\begin{eqnarray} 
  P^{\rm eq}_N &=& \frac{{\rm tr}_N \exp(-\beta H^D)}{{\rm tr}
  \exp(-\beta H^D)} =\frac{{\rm tr}_N \exp(-\beta H^D)}{\sum_{N} {\rm
  tr}_N \exp(-\beta H^D)} \ .
\label{pndot}
\end{eqnarray}
The inverse temperature is denoted by $\beta$, and ${\rm tr}_N$
denotes the trace over the Fock states with $N$ electrons on the
dot. The canonical occupation number of level $n$ when there are $N$
electrons on the dot is given by 
\begin{eqnarray}
F^{\rm eq}(\epsilon_i|N) &\equiv& \langle \hat{n}_i \rangle_N = 
\frac{{\rm tr}_N ( n_i \exp(-\beta H^D))}{{\rm tr}_N \exp(-\beta H^D)}
\label{occun} \ .
\end{eqnarray}
Eqn. (\ref{condbeen}) has also been derived \cite{lee} using a
Landauer--B\"uttiker type approach generalized to include the
interaction of electrons on the dot.

Numerically, it turns out to be sufficient to calculate $F^{\rm
  eq}(\epsilon_n|N)$ for a window of levels around the Fermi energy,
and to take the occupation numbers equal to 1 below and 0 above this
window. The following results are obtained using a window of 16 levels
that are populated with 8 electrons. We checked that our results are
insensitive to changes of the window size. 

Using Eq.~(\ref{tunnel}) and the relation $\Gamma^{L(R)}_n\sim
\sum_{k} |V_{k,n}^{L(R)}|^2 \delta (\epsilon_n-\epsilon_k^{L(R)})$, we
find that the tunneling rates $\Gamma_n$ are proportional to the
square modulus of the harmonic oscillator wave functions at the
position of the barriers $\vec{r}=(x_B, y_B)$,  
\begin{eqnarray}
  \Gamma_n &\sim& \frac{(H_{n_x}(\xi_x) H_{n_y}(\xi_y))^2}{2^{(n_x+n_y)}
  \ n_x! \ n_y!} e^{-(\xi_x^2 + \xi_y^2)} \ ,
\label{modelrates}
\end{eqnarray}
where $H_n$ are the Hermite polynomials and where we have used
dimensionless variables $\xi_x=\sqrt{m\omega_x/\hbar} \ x_B$ and
$\xi_y=\sqrt{m\omega_y/\hbar} \ y_B$. Neglecting deformation we can
relate the barrier height $V_B$ to the position of the barrier via
$V_B/\hbar \omega_x=(\xi_x^2+\xi_y^2)/2$. In the sequel we choose
a fixed value $V_B/\hbar \omega=25$ for the barrier height.
\footnote{Compared with typical experimental parameters this value
  appears to be too small. We use this value in order to avoid an
  unphysically large increase of $\Gamma$ with increasing quantum
  number $n_x$. Such a strong increase is characteristic of the
  harmonically shaped barrier. A less steep increase would be obtained
  for steeper tunneling barriers. Such barriers appear to give a more
  realistic description of the depletion zone of the electron gas near
  the gates. For reasons of consistency with the harmonic oscillator
  model used throughout this paper, we decided to use harmonic
  barriers.} We consider two different geometries of the leads
connecting the quantum dot with external reservoirs: (i) The two leads
are located exactly opposite to each other, so that $\xi_x = \pm
\sqrt{50}$ and  $\xi_y=0$. (ii) The leads are arranged at an angle of
90 degrees, with the barriers at $\xi_x=\xi_y=-\sqrt{50/2}$ and $\xi_x
= -\xi_y = \sqrt{50/2}$. The latter geometry has been used, for
instance, in the experiment of Ref.~\cite{marcus}.

The tunneling rates $\Gamma_n$ for both geometries and for several
states $n$ are presented in Table~\ref{tabrates}. We note that for
geometry (i) the flat levels are coupled much more strongly to the
leads than the steep ones (columns 1 and 2). Comparing the rates for
the flat levels of different shells we find a considerable increase
with increasing $n_x$ (columns 4 and 1). For geometry (ii) the
states within one shell with equal quantum numbers $n_x$ and $n_y$
are most strongly coupled to the leads. However, the difference
between strongly and weakly coupled states is not as pronounced as in
geometry (i).

\begin{figure}[t]
\centerline{\psfig{file=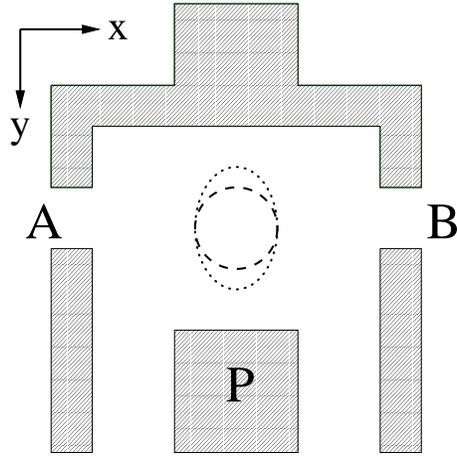,width=6cm,height=6cm,angle=270}}
  \caption[]{
    A quantum dot (schematic). Electrons enter or
    leave the dot through the tunneling barriers $A$ and $B$. The
    plunger gate $P$ controls the number of electrons on the
    dot. Equipotential lines of the confining potential are shown for
    the isotropic (dashed) and deformed (dotted) case.} 
  \label{geom}
\end{figure}

\begin{figure}[b]
\centerline{\psfig {file=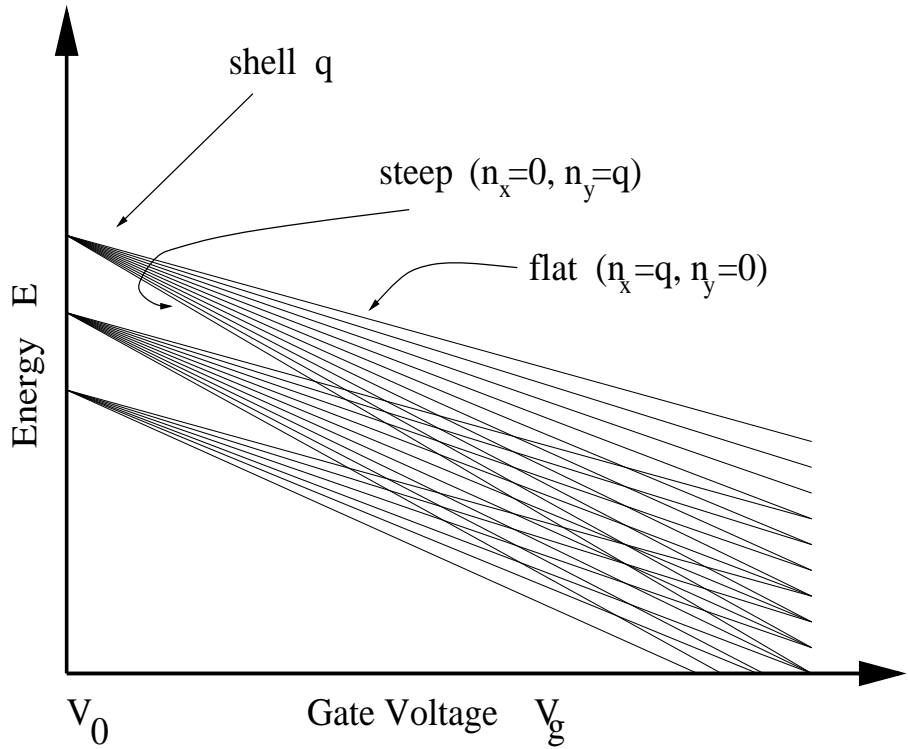,width=12cm,height=10cm,angle=270}}
  \caption[]{
    Dependence of the single--particle energies of the dot on the
    gate voltage $V_g$.}
  \label{shell}
\end{figure}

\begin{figure}[b]
\centerline{\psfig{file=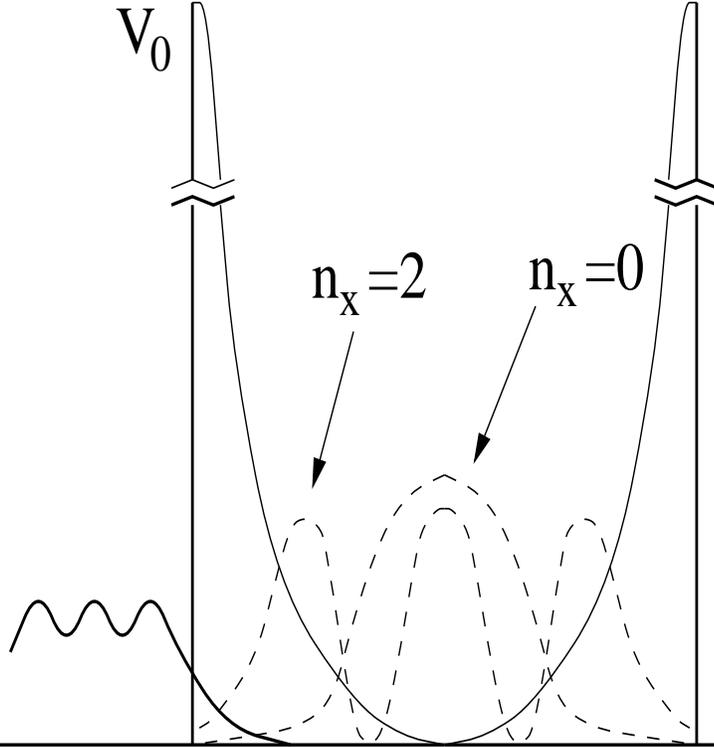,width=12cm,height=10cm,angle=270}}
  \caption[]{The thin solid line shows a cross section of the
    potential in longitudinal direction, the two barriers lying
    at opposite ends. The overlap of the dot wave functions (probability
    shown as dashed lines) and of the lead wave function (probability
    shown as a solid line on the left) increases strongly with the
    quantum number $n_x$ in x--direction.}
\label{over}
\end{figure}

\begin{figure}[b]
\centerline{\psfig {file=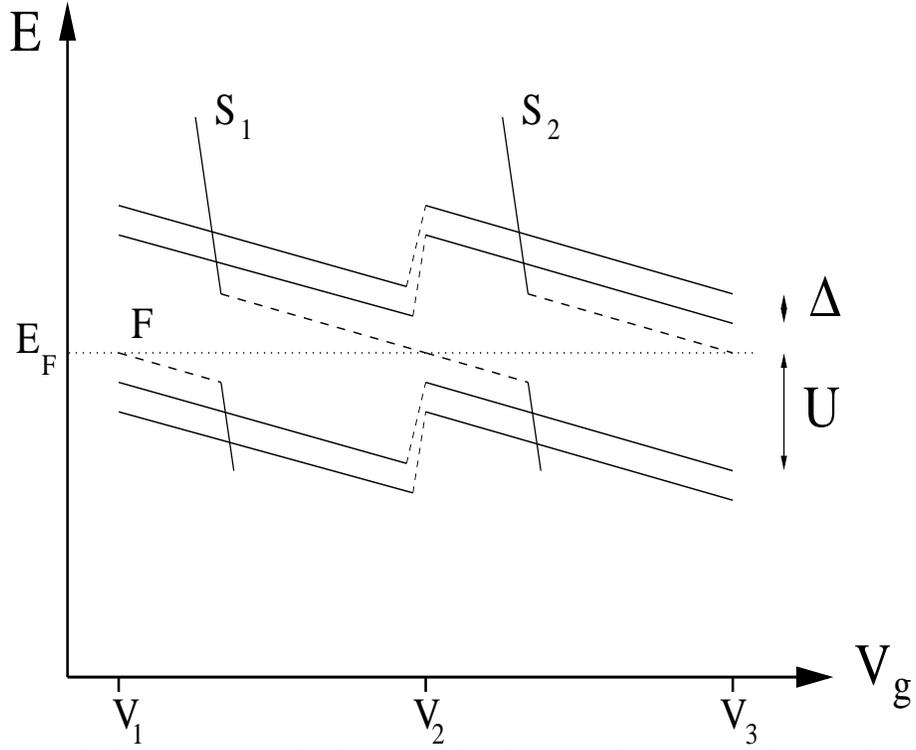,width=12cm,height=10cm,angle=270}}
  \caption[]{
    Avoided crossings with the steep levels $S_1$ and $S_2$ cause the
    flat level $F$ (dashed) to stay close to the Fermi energy
    $E_F$. Resonances dominated by this level occur at gate voltages
    $V_1,V_2,V_3$. There is a gap of magnitude $U$ between the last
    occupied and the first empty level.} 
  \label{avoid}
\end{figure}

We conclude that {\em independently of the precise shape of the
  barrier}, in a geometry of type (i) which is realized in
Figure~\ref{geom} flat levels are more strongly coupled to the leads
than steep levels. This is because the wave functions of flat levels
have cigar--like shapes extending closer to the leads (cf. 
Figure~\ref{over}). Therefore, flat levels carry the bulk of the
current. The difference between well coupled and poorly coupled dot
states is less pronounced when the leads are arranged at an angle
(geometry (ii)). In this case, the mechanism for peak correlations
described in Sec.~\ref{Model} leads us to expect a reduction in the
length of sequences of correlated peaks.

We present results for $\alpha=1$, $\gamma=0.005$, $E_0=-11$ and
$V_0=90$. At zero deformation, the mean level spacing $\Delta$ is
related to the harmonic oscillator frequency $\omega_x$ in
$x$-direction by
\begin{eqnarray}
\Delta=\frac{E_F}{N_{el}}&=&\frac{\hbar \omega_x
  N_{sh}}{\frac{1}{2}(N_{sh}+1)(N_{sh}+2)}\approx \frac{2 \hbar
  \omega_x}{N_{sh}} \ .  
\end{eqnarray}
Here $N_{sh}$ is the number of the last filled shell and $N_{el} =
(N_{sh}+1) (N_{sh}+2) / 2$ is the total number of electrons on the
dot. We take $N_{sh}=14$ so that there are about 100 electrons on the
dot. We assume $\Delta=0.03 U$ which roughly corresponds to the
situation of the experiments of Refs.~\cite{marcus,heiblum}. 

\begin{figure}[h!]
\centerline{\psfig{file=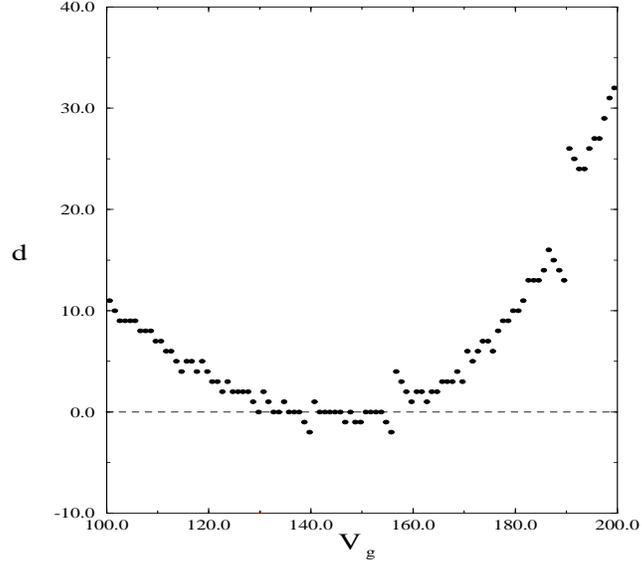,height=9cm,width=9cm,angle=0}}
\caption[]{Distance $d$ (in number of levels) of the flat level
  $(n_x=14, n_y=0)$ from the Fermi energy $E_F$ versus gate voltage
  $V_g$.}
\label{distant}
\end{figure}

In order to monitor the influence of a flat level $F$ on the
conductance, we define the distance $d$ of $F$ from the Fermi energy
as the {\it number} of levels between $F$ and the Fermi energy
including $F$ itself and count positively (negatively) for states
above (below) the Fermi energy. Thus, $d=1$ indicates that $F$ is the first
unoccupied level. 
Figure~\ref{distant} shows $d$ vs. gate voltage for the flat
level $n_x=14, n_y=0$. Adjacent points correspond to adjacent Coulomb 
blockade resonances. For $130<V_g<160$, $F$ stays in the vicinity of
the Fermi energy. Hence there is -- on average -- one crossing with a
steep level in the interval $\delta V_g$. For $V_g<130$ ($V_g>160$),
the number of avoided crossings in the interval $\delta V_g$ is less
than one (bigger than one), and $F$ moves towards the (away from the)
Fermi energy, respectively. The jumps at $V_g=156$ and $V_g=180$
are due to multiple crossings of levels which occur because of the
integrability of our model.

\begin{figure}[h!]
\begin{minipage}{7.5cm}
\centerline{\psfig{file=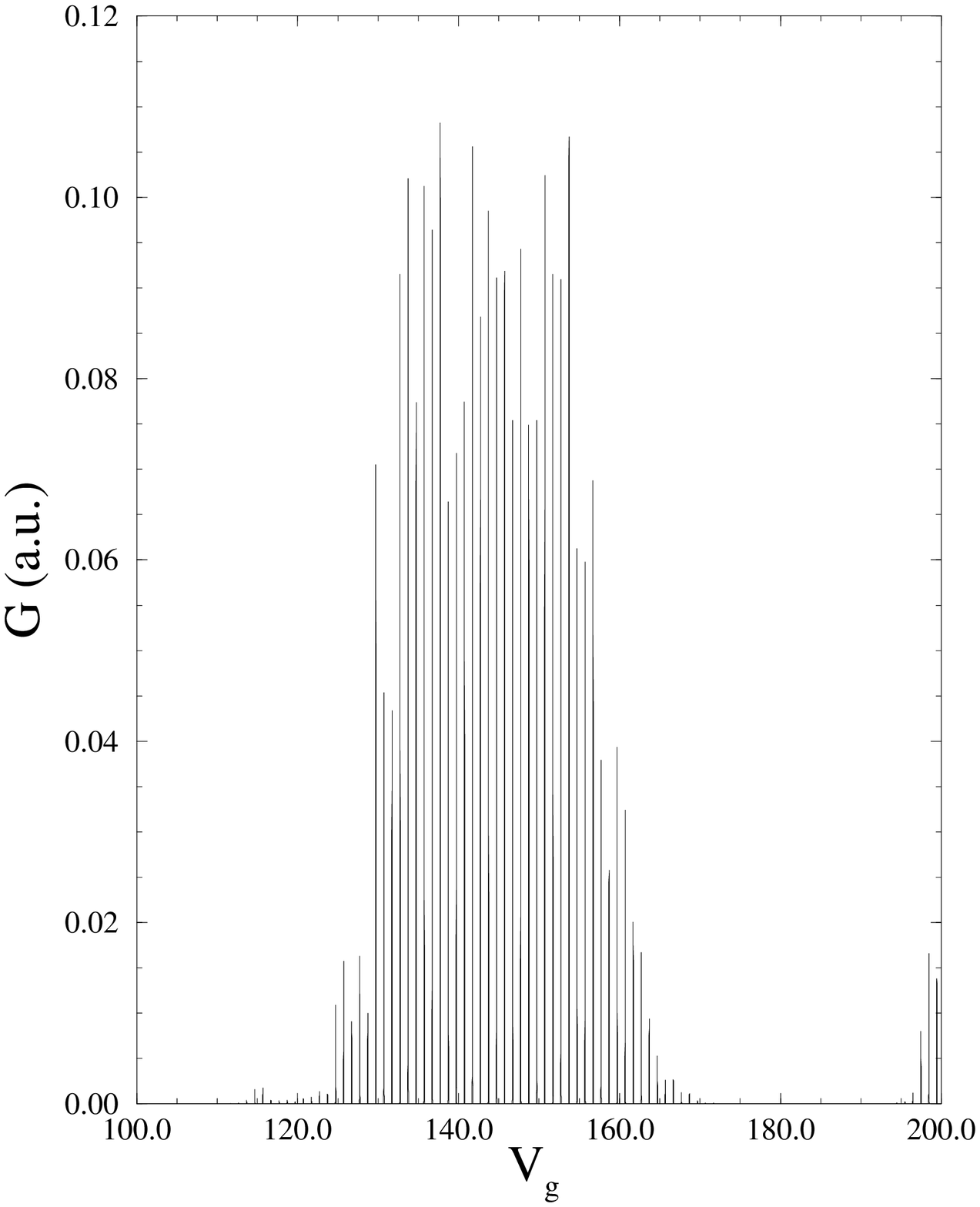,height=7cm,width=7cm,angle=0}}
\end{minipage}
\begin{minipage}{7.5cm}
\centerline{\psfig{file=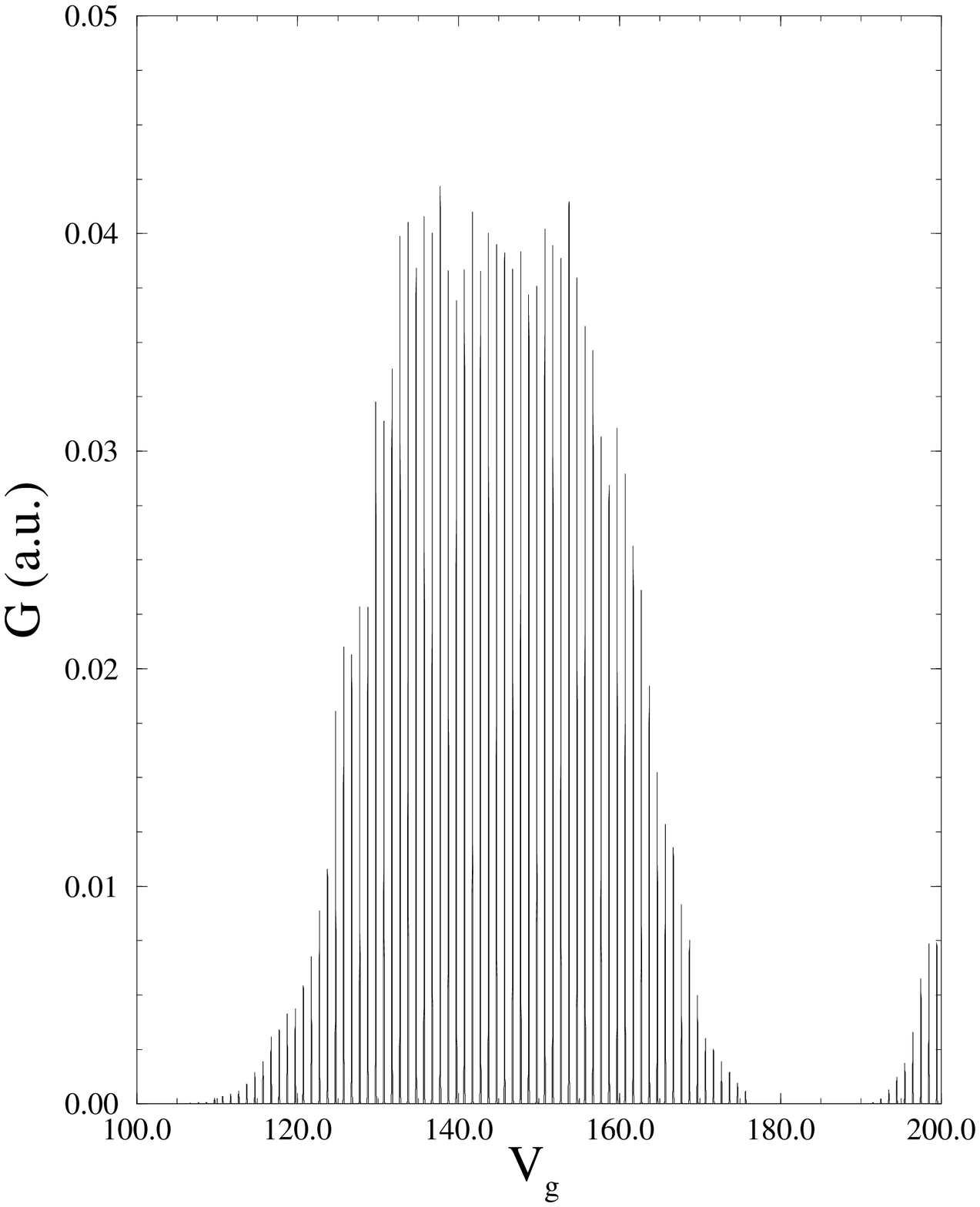,height=7cm,width=7cm,angle=0}}
\end{minipage}
\caption{Conductance $G$ vs. gate voltage for (a) $kT=\Delta/5$ (left)
  and (b) $kT=2\Delta/5$ (right).}
\label{conduct}
\end{figure}

For the same choice of parameters as in Figure~\ref{distant} and for
geometry (i), Figure~\ref{conduct} shows the conductance vs. gate
voltage for two temperatures, (a) $kT=0.2 \Delta$ and (b) $kT=0.4
\Delta$. About 100 Coulomb blockade resonances occur in the interval
$100 < V_g < 200$. In both plots strong peaks with similar peak
heights appear whenever the flat level $F$ is close to the Fermi
energy, especially at the higher temperature (case (b) of
Figure~\ref{conduct}). In the regions $V_g < 130$ and $V_g > 160$
where the conductance is not dominated by the flat level $F$, $G$ is
much smaller than in the interval $130<V_g<160$. On the scale of
Fig.~\ref{conduct}, some of the conductance peaks are not even
visible. 

\begin{figure}[h!]
\centerline{\psfig{file=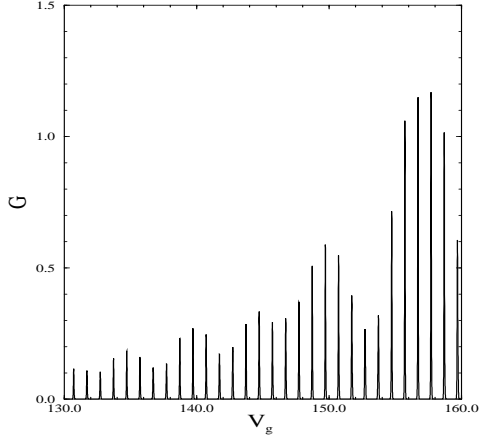,height=7cm,width=7cm,angle=0}}
\caption[]{Conductance $G$ vs. gate voltage $V_g$ for the same
  parameters as in Figure~\protect{\ref{conduct}} (b) but with the
  leads arranged at an angle of 90 degrees.}
\label{ninty}
\end{figure}

Figure~\ref{ninty} shows the conductance vs. gate voltage for the same
parameters as in case (b) of Figure~\ref{conduct} but for geometry
(ii). Now, the resonances are dominated by steeper levels from higher
shells which are coupled more strongly to the leads than the flat
level. This is why the peak heights are bigger on average than in
Figure~\ref{conduct}, why they show stronger variation, and why they
increase systematically with increasing gate voltage.

\section{Phase}
\label{Phase}

We now turn to the behavior of the phase of the transmission amplitude
through the quantum dot. This phase has recently been measured in a
set of experiments using an Aharonov--Bohm (AB) interferometer with a
quantum dot embedded in one of its arms. We consider the simplest case
where the AB interferometer is coupled to only one channel in each
connecting lead. The transmission coefficient $\cal T$ through the
AB device is then given by
\begin{eqnarray}
{\cal T} \approx {\cal T}_0 + 2 {\rm Re} \left\{  t_0^*e^{-2 \pi i
  \Phi/\Phi_0} \int dE 
  \left( -\frac{\partial f}{\partial E} \right) t_{QD}(E) \right\} \ . 
\label{tau}
\end{eqnarray}
Here, ${\cal T}_0 = |t_0|^2$ is a flux-- and energy--independent term
given by the square of the amplitude for transmission through the
empty arm of the AB interferometer, while $t_{QD}$ is the amplitude
through the arm containing the dot. Since the quantum dot is weakly
coupled to the arm, we have $|t_{QD}| \ll |t_0|$. We have explicitly
displayed the dependence on the magnetic flux $\Phi$ through the AB
device and neglected higher harmonics. The symbol $\Phi_0=h/e$ denotes
the elementary flux quantum.

The master equation used in Section \ref{CondOsc} deals with occupation
probabilities and is, therefore, not able to yield the phase of the
transmission amplitude $t_{QD}$. We have used another approach. We have
expressed $t_{QD}$ in terms of the retarded Green function $G^{\rm
  ret}$ of the dot,
\begin{eqnarray}
t_{QD}(E)=\sum_{i,j} V_i^L(E) G^{\rm ret}_{ij}(E) V_j^{R*}(E) \ .
\end{eqnarray}
The finite--temperature Green function $G^{ret}$ must be calculated in
the presence of the interaction $U$ and the tunneling. A derivation of
$G^{ret}$ starting from the equations of motion is given in Appendix
\ref{eqm}. Assuming that the total number of electrons on the dot is a
constant of motion, we obtain
\begin{eqnarray}
  G_{ij}^{\rm ret}(E)&\approx & \delta_{ij}\sum_{N=0}^{\infty}
  P^{\rm eq}_N \left[ \frac{1-\langle \hat{n}_i
    \rangle_N}{E-(\epsilon_i-\mu+U N)+i \Gamma_i/2} +
  \frac{\langle \hat{n}_i \rangle_N}{E-(\epsilon_i-\mu+U
    (N-1))+i\Gamma_i/2}\right] \ .
\label{Gret}
\end{eqnarray}
Here $\Gamma_i=\Gamma_i^L+\Gamma_i^R$.
The probability $P^{\rm eq}_N$ that there are $N$ electrons on the
dot and the canonical occupation number $\langle \hat{n} \rangle_N$
are given in Eqs.~(\ref{pndot}) and (\ref{occun}), respectively.

Within our approximations the Green function $G^{ret}$ is diagonal. 
This fact implies that real (\ particle--hole) excitations of the dot
caused by tunneling transitions are not taken into account. This is 
justified in the regime of elastic cotunneling $kT < \sqrt{U \Delta}$
where inelastic cotunneling processes do not contribute significantly
to the transmission \cite{avnaz}. Equation~(\ref{Gret}) is a good
approximation to 
the exact retarded Green function between Coulomb blockade resonances
where fluctuations in the occupation number of the dot are strongly
suppressed. Moreover, {\em even at resonance} where $G^{ret}$ reduces
to a single Breit--Wigner term, Eq.~(\ref{Gret}) is expected
\cite{Stafford} to apply provided there are no degeneracies and we
work well above the Kondo temperature \cite{kondo}. The success of
Eq.~(\ref{Gret}) in these limiting cases suggests that well above the
Kondo temperature, Eq.~(\ref{Gret}) is a good approximation to the
exact Green function for all energies.

Combining Eqs.~(\ref{Gret}) and (\ref{tau}) we obtain
\begin{equation}
{\cal T} = {\cal T}_0 + {\rm Re}\  t_0^*\  \widetilde{t_{QD}}\  e^{-2\pi i
\Phi/\Phi_0}, 
\end{equation}
where
\begin{eqnarray}
\widetilde{t_{QD}} &=& \int dE 
  \left( -\frac{\partial f}{\partial E} \right) t_{QD}(E) \nonumber \\
   &=& \frac{\beta}{2\pi i} \sum_i \sum_{N=0}^{\infty} V_i^L V_i^{R*}
  P^{\rm eq}_N \left[ (1-\langle \hat{n}_i \rangle) \  
  \psi^{(2)}\left( \frac{\beta}{2\pi i} (i\Gamma_i
  - \epsilon_i + \mu - UN) +\frac{1}{2} \right) \right. \nonumber \\
 &\mbox{}& \left. + \langle \hat{n}_i \rangle \ 
  \psi^{(2)}\left( \frac{\beta}{2\pi i}(i\Gamma_i -\epsilon_i  +\mu -
  U (N-1) ) + \frac{1}{2}\right)  \right]
\end{eqnarray}
with the trigamma function $\psi^{(2)}$.

\begin{figure}[h!]
\begin{minipage}{7.5cm}
\centerline{\psfig{file=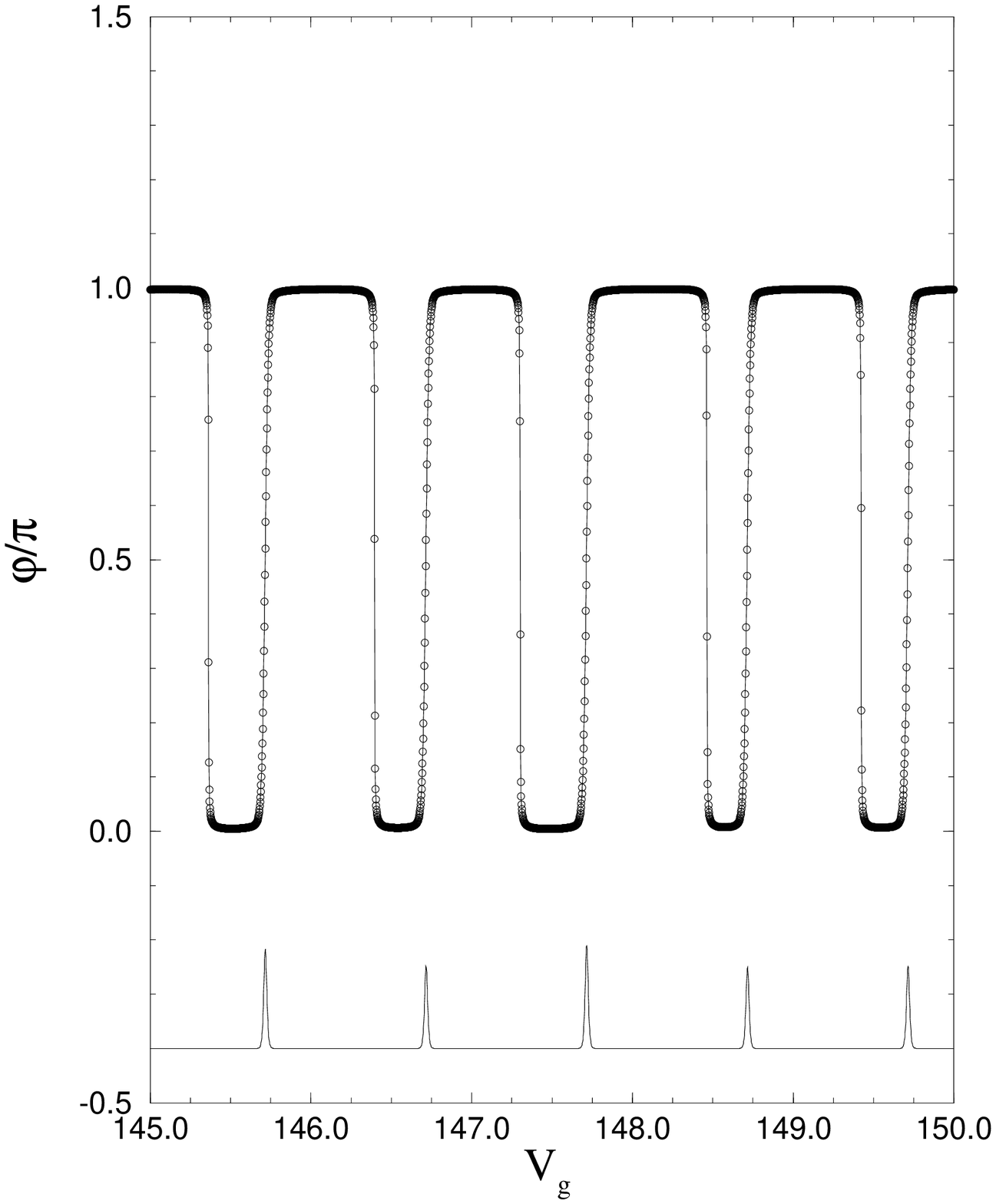,height=7cm,width=7cm,angle=0}}
\end{minipage}
\begin{minipage}{7.5cm}
\centerline{\psfig{file=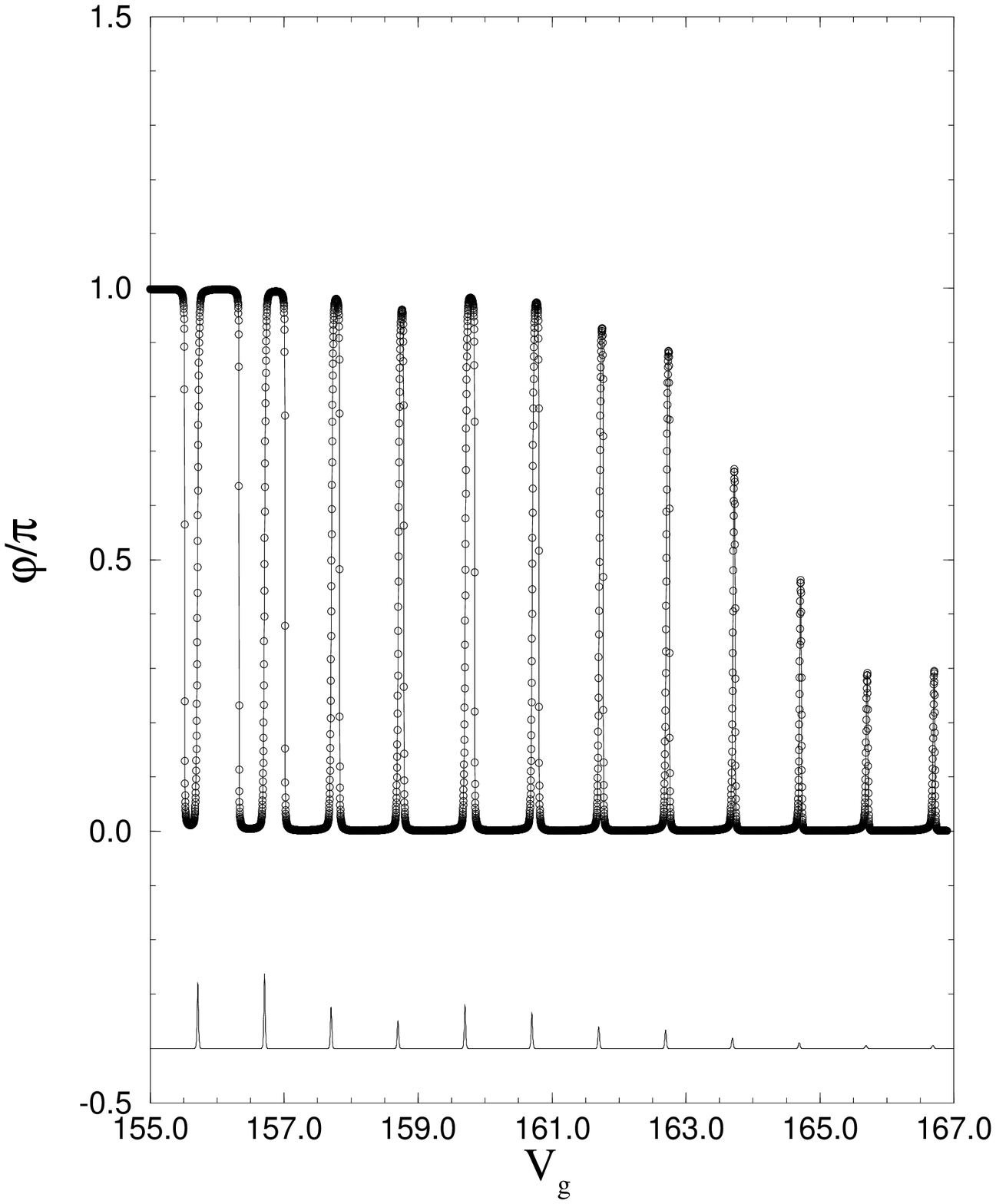,height=7cm,width=7cm,angle=0}}
\end{minipage}
\caption{Phase $\phi$ of the transmission amplitude versus gate
  voltage $V_g$ at $kT=\Delta/5$ in two different intervals.
  The solid lines at the bottom of the plots display the conductance
  peaks. In the case shown in the left (right) part, the flat level
  $n_x=14, n_y=0$ is at or near (far removed from) the Fermi level,
  respectively. In the case of the right part, the flat level
  influences the phase as a background only.}
\label{phase}
\end{figure}

\begin{figure}[h!]
\centerline{\psfig{file=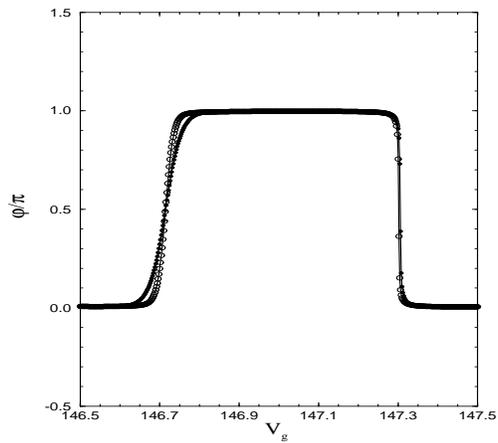,height=7cm,width=7cm,angle=0}}
\caption[]{Transmission phase for $kT=0.2\Delta$ (open circles) and 
$kT=0.4 \Delta$ (filled circle). The increase by $\pi$ at the
resonance takes place on the scale $kT$, the phase lapse between
resonances, on the scale $\Gamma$.} 
\label{tempphase}
\end{figure}

In Figure~\ref{phase} we show the phase $\phi$ of the transmission
amplitude versus gate voltage. As in the calculation of the conductance 
in Section~\ref{CondOsc}, the canonical occupation numbers are obtained 
by distributing 8 electrons over a window containing 16 levels. We take 
$\Gamma_i=\Gamma=0.002U= \Delta/15$. The solid lines at the bottom of
the plots show the conductance peaks and help to identify the
resonance positions. In the left part of Figure~\ref{phase} the flat
level $n_x=14$, $n_y=0$ is close to the Fermi energy. Here we find a
strikingly similar behavior of the phase at all resonances. This
behavior is found not only within the $V_g$ interval shown but {\em
  for the entire interval $130 < V_g < 160$ comprising 30
  resonances}. The phase regularly increases by $\pi$ at resonance and
displays a sharp lapse by $\pi$ between adjacent resonances. As
observed in Refs.~\cite{yy,hw} the increase at resonance occurs on the
scale $kT$ (we assume $kT>\Gamma$) and the phase lapse between
resonances on the scale $\Gamma$. The temperature dependence of the
phase is shown in Figure~\ref{tempphase}. In the right part of
Figure~\ref{phase} we show the transmission phase for the case where
the distance between the flat level and the Fermi energy is large
compared to $kT$ and increases with $V_g$ (cf.\ Figure~\ref{distant}). 
The phase behaves less regularly, with an increase by $\pi$ or less at
and an immediate phase lapse near the resonances. Between resonances
the phase remains virtually constant.

To interpret our results, we consider first the phase $\phi$ at
resonance. The identical behavior of $\phi$ at all resonances in
Figure~\ref{phase}(a) reflects the fact that at each resonance, the
transmission through the dot is dominated by the strongly coupled
level $F$. This is the same mechanism as in the sequence of strong
conductance peaks shown in Figure~\ref{conduct}(a). The more erratic
phase behavior seen in Figure~\ref{phase}(b) is the result of the
interplay of various levels of the dot. The regular behavior of the
phase lapse between adjacent resonances is also due to the dominance
of the flat level. 
At finite temperature the flat level $F$
has a finite probability of being either occupied or empty and, thus,
may contribute to both an electron--like and a hole--like cotunneling
process. The contribution of both processes to the transmission
amplitude through the dot is
\begin{eqnarray}
  t_F& = & V_F^L V_F^{R*} \left[ \frac{1-\langle \hat{n}_F
    \rangle_N}{E-(\epsilon_F-\mu+U N)+i \Gamma_F/2} +
  \frac{\langle \hat{n}_F \rangle_N}{E-(\epsilon_F-\mu+U
    (N-1))+i\Gamma_F/2}\right]
\label{cotun}
\end{eqnarray}
where the first (second) term represents the electron--like (hole--like)
contribution, respectively. As the gate voltage $V_g = \mu / \alpha$
scans the $N^{\rm th}$ valley (i.e. varies from $(\epsilon_N + U\cdot
(N-1))/\alpha$ to $(\epsilon_{N+1} + U \cdot N)/\alpha$), the sign of
$Re \ t_F$ reverses, leading to a phase lapse. The same conclusion has
previously been reached in Ref.~\cite{yy}; an interpretation in terms
of scattering theory has been given in Ref.~\cite{hw}. If $F$ is far
away from the Fermi energy (on the scale of $kT$) either the
particle--like or the hole--like process will dominate, and the phase
lapse moves from the valley towards the resonance, as depicted in
Figure \ref{phase}(b).

We emphasize that the phase lapse between resonances is a genuine
interaction effect. Indeed, the interaction $U$ is needed to keep the
flat level close to the Fermi energy for a long sequence of resonances. 
For non--interacting particles ($U \to 0$) the transmission amplitude
at different resonances would be dominated by different
single--particle levels. In the same limit, the cotunneling amplitude
(\ref{cotun}) would reduce to a single, temperature--independent
term. The phases of the transmission amplitude in consecutive valleys
would not be correlated, and there would be no systematic phase lapse
between resonances. We also note that the systematic phase lapse
occurs only at finite temperature. At zero temperature a flat level
could only contribute to either particle--like or hole--like 
cotunneling. 

\section{Summary. The Question of Non-Universality}

Since the first measurements on quantum dots in Aharonov-Bohm
interference devices were reported, many aspects of phase--coherent
transport through quantum dots have been understood theoretically. 
However, one of the most striking features, the strong correlations
of the transmission phases in sequences of many resonances, has long
withstood a satisfactory theoretical explanation. Earlier attempts
\cite{hhw,yy} to solve the problem could account for short sequences
but not for the sequences of more than 10 resonances found
experimentally.

In this paper we have demonstrated the viability of a mechanism, based
on a synthesis of the ideas proposed in Refs.~\cite{hhw,yy}, that
gives rise to long sequences of correlated peak heights and
transmission phases. We have used several approximations, the most
central one being that the confining potential defining the dot is
``almost'' integrable. More precisely, both the deviation from
integrability and the disorder must constitute a perturbation which is
small on the scale of the mean single--particle level spacing. 
Our model shares many features of the ballistic quantum
dots employed in some experiments \cite{marcus,heiblum} and may
account semiquantitatively for some of the experimental observations.
Nevertheless, our
analysis naturally falls short of providing a complete and universal
framework which could account for the combined effect of disorder and
interaction on correlations in transmission experiments.

We have used the following specific conditions and assumptions: (i)
Among the eigenstates of the quantum dot, some must be coupled more
strongly to the leads than others. This assumption is met by a model
which is nearly integrable, as is the case for our parabolic confining
potential. This potential renders $(n_x, n_y)$ good quantum numbers
for all values of $V_g$. Of all levels in a shell, the level with $n_y
= 0$ is most strongly coupled to the leads. (ii) Changes in the gate
voltage induce deformations of the dot boundary such that the
potential is deformed in the transverse $y$--direction. This
assumption guarantees that in each shell, the level with $n_y = 0$ is
flat, i.e., stays close to the Fermi energy over a wide range of
$V_g$--values. (iii) On average there is one crossing of the flat
level with one other level per unit interval. 
This interval is defined by the change in $V_g$ needed to add an
extra electron to the dot. (iv) The temperature is sufficiently high
to produce sufficiently long sequences of correlated resonances. The
minimum temperature required by this condition depends both on the
distance of the most strongly coupled level from the Fermi energy, and
on the relative strength of the coupling of that level to the
leads. With increasing temperature the correlations become more robust.

Strong boundary deformations leading to strongly chaotic classical
motion within the dot, or strong disorder in the dot are likely to
destroy the correlations altogether since they generically do not
allow for the existence of eigenstates that are particularly well
coupled to external leads. In this sense, the correlations proposed in
the present paper are {\em non-universal in origin}. This conclusion
agrees with the observation of weak peak correlations in the strongly
deformed dots of Ref.~\cite{marcus} as compared with the strong
correlations found in the experiments of Ref.~\cite{heiblum}.

Our ideas may be checked experimentally on dots that are not embedded
in an AB device but are coupled directly to leads. This setup does not
allow for tests of phase correlations but provides a convenient setup
for measuring conductance peak correlations. The conductance of a dot
with a regular (rectangular) lithographic shape has been measured by
Simmel et {\it al.} \cite{wharam}. These authors did indeed find a
sequence of more than 10 strong peaks with very similar peak
heights. In the same sweep they also observe envelopes of smaller
peaks very similar to our results in Figure~\ref{conduct}. To test our
picture further, it would be illuminating to perform a similar
two--terminal conductance experiment with leads attached at two sides
of the dot which form an angle of 90 degrees. Here, a reduction of the
correlations is to be expected. It would also be interesting to
compare two setups, one with a plunger gate and the other with a
backgate configuration. In the latter case the potential deformation 
is reduced. This should suppress our correlation mechanism.

\section{Acknowledgment}
This work was supported by the German-Israeli Foundation (GIF), by the
Israel Science Foundation founded by the Israel Academy of Sciences and
Humanities-Centers of Excellence Program, and by the U.S.-Israel
Binational Science Foundation (BSF).

\begin{appendix}
\section{Retarded Green Function}
\label{eqm}
We derive Eq.~(\ref{Gret}). The retarded Green function is defined as
\begin{eqnarray}
  G^{ret}_{kl}=&-i\theta(t)\ \langle [c_k(t),c_l^{\dagger}(0)]_+
  \rangle\ &=
  F_{kl}+G_{kl}\ \ \ \ \mbox{where}\\
  G_{kl}=&-i\theta(t) \langle c_k(t) c_l^{\dagger}(0) \rangle\ \,
  \ \ \ \ &F_{kl}= -i\theta(t) \langle c_l^{\dagger}(0) c_k(t)
  \rangle.
\end{eqnarray}
The brackets denote the thermal average, $\langle ... \rangle= {\rm
  tr}(... \exp(-\beta H))/tr (\exp(-\beta H))$. With $P_N=tr_N
  \exp(-\beta H)/tr \exp(-\beta H)$, we write the trace as a sum of
terms with a fixed number $N$ of electrons on the dot,
\begin{eqnarray}
  G_{kl}&=& -i\theta(t) \sum_{N=0}^{\infty} P_N \langle c_k(t)
  c_l^{\dagger}(0) \rangle_N= \sum_{N=0}^{\infty} G_{kl}^{(N)} \ .
\end{eqnarray}
In the ``equation of motion'' method the Green function is
differentiated with respect to time. Since the time evolution of an
operator is given by the commutator with the Hamiltonian, a system of
differential equations containing higher--order Green functions is
generated. A closed system is obtained if these Green functions can be
approximately uncoupled. The solution is obtained by Fourier
transformation. Specifically,
\begin{eqnarray}
  \frac{\partial}{\partial t} G_{ij}^{(N)} &=& -i \delta(t)
  \langle c_i(0) c_j^{\dagger}(0) \rangle_N -\theta(t) \langle [c_i,
  H](t) c_j^{\dagger}(0) \rangle_N \\
  \mbox{} &=& -i \delta(t) \langle c_i(0) c_j^{\dagger}(0) \rangle_N -
  \theta(t) \left( (\epsilon_i-\mu) \langle c_i(t) c_j^{\dagger}(0)
  \rangle_N + U \langle N(t) c_i(t) c_j^{\dagger}(0) \rangle_N \right)
  \nonumber\\
  &\mbox{}& -\theta(t) \sum_{k} \left( V_{ki}^{L*} \langle a_k(t)^L
  c_j^{\dagger}(0) \rangle_N + V_{ki}^{R*} \langle a_k(t)^R
  c_j^{\dagger}(0) \rangle_N \right). 
\label{eom}
\end{eqnarray} 
Since the interaction contains two creation and two annihilation
operators a two particle Green function appears in the second line.  
The last two terms stem from the coupling to the leads, and by another
equation of motion can be expressed in terms of the Green function for
the dot. Assuming that the states in the lead and in the dot are
uncorrelated at $t=0$ we obtain
\begin{eqnarray}
\langle a_k(t)
  c_j^{\dagger}(0) \rangle_N &=& \sum_{i} V_{ki}^{L} \int d\bar{t}
  G_{ij}^{(N)}(t-\bar{t}) \theta(\bar{t}) \exp(-i\epsilon_k^{L}
  \bar{t}) \ .
\end{eqnarray}
Fourier transformation of Eq.~(\ref{eom}) yields
\begin{eqnarray}
  \omega G_{ij}^{(N)} &=& \langle 1-n_i \rangle_N \delta_{ij} + 
  (\epsilon_i -\mu) G_{ij}^{(N)}+ U \tilde{G_{ij}}+\nonumber \\
&\mbox{}&+
  \sum_{kl} \left( \frac{V_{ki}^{L*} V_{kl}^L}{\omega
  -\epsilon_k^L+i \delta} +\frac{V_{ki}^{R*} V_{kl}^R}{\omega
  -\epsilon_k^R+i \delta} \right) G_{lj}^{(N)}
\end{eqnarray}
where $\tilde{G_{ij}}=-i \int dt \theta(t) \langle N(t) c_i(t) c_j^{\dagger}
\rangle_N \exp i\omega t$. The equations of motion are closed by
assuming the total occupation $N$ to be constant. Then the number 
operator can be taken out of the expectation value. The assumption is
justified in the valleys between resonances whereas at each resonance,
$N$ fluctuates. For isolated resonances (level width $\ll$ level
spacing), we have Im $\sum_{k} V_{ki}^{L*} V_{kl}^{L}/(\omega -
\epsilon_k^{L} + i\delta) = -i\delta_{il} \Gamma_i^L/2$. This yields
\begin{eqnarray}
  G_{ij}^{(N)} &=& \frac{\delta_{ij} \langle 1-n_i \rangle_N} {\omega
  -(\epsilon_i -\mu+U N)+ i \Gamma_i^{L}/2+ i\Gamma_i^{R}/2} \ .
\end{eqnarray}
For $F_{ij}$ we proceed analogously and eventually obtain
Eq.~(\ref{Gret}). 
\end{appendix}

\begin{table}
\begin{tabular}{|c||c|c|c|c|c|} 
\hline
$n= (n_x, n_y)$ & (14,0) & (0,14) & (6,8) & (12,0) & (6,6) \\ \hline \hline
$\Gamma_n$ for geometry (i) & $2.68 \cdot 10^{-6}$ & $4.04 \cdot
10^{-23}$ & $5.34 
\cdot 10^{-14}$ & $9.09 \cdot 10^{-8}$ & $6.11 \cdot 10^{-14}$ \\
\hline
$\Gamma_n$ for geometry (ii) & $6.30 \cdot 10^{-12}$ & $6.30 \cdot
10^{-12}$ & $5.66 
\cdot 10^{-7}$ & $2.88 \cdot 10^{-12}$ & $2.41 \cdot 10^{-8}$ \\
\hline
\end{tabular}

\caption[]{Tunneling rates for several levels $n$ calculated from
  Eq.~(\protect{\ref{modelrates}}). In geometry (i) the leads are
  opposite to each other with $\xi_x=\sqrt{50}$ and $\xi_y=0$. In
  geometry (ii), the leads are arranged at an angle, with
  $\xi_x=\xi_y=\sqrt{50/2}$.}
\label{tabrates}
\end{table}

\newpage

\newpage

\newpage

\newpage

\end{document}